%% file: MBS-PopDynPluv.tex
\documentclass[preprint,11pt]{elsarticle}

\usepackage{graphicx}
\usepackage{epstopdf}
\usepackage{amssymb, amsmath}
\usepackage{tikz}
\usepackage{slashbox}
\usepackage{hyperref}
\hypersetup{
     colorlinks   = true,
     citecolor    = blue,
     colorlinks=true,
     linkcolor=blue,
     urlcolor=blue
}

\journal{Mathematical Bioscience}

\newcommand{\scaler}[2]{ \resizebox{!}{#2}{#1} }
\newtheorem{thm}{Theorem}
\newcounter{rem1}
\newcounter{prf1}
\newcounter{def1}

\newtheorem{rem}[rem1]{Remark}
\newtheorem{prf}[prf1]{Proof}
\newtheorem{defn}[def1]{Definition}

\begin{document}
\begin{frontmatter}


  \title{A model to predict the population size of the dengue fever vector based on rainfall data} 

\author[cefetmg]{L. S. Barsante}
\author[icbufmg]{K. S. Paix\~ao}
\author[icbufmg]{K. H. Laass}
\author[cefetmg]{R. T. N. Cardoso}
\author[icbufmg]{\newline \'{A}. E. Eiras}
\author[cefetmg]{J. L. Acebal \corref{cor1}}
\ead{acebal@dppg.cefetmg.br}

\cortext[cor1]{Corresponding author: acebal@dppg.cefetmg.br}

\address[icbufmg]{Laborat\'orio. Ecologia Qu\'\i mica de Insetos Vetores, Instituto de Ci\^encias Biol\'ogicas, Universidade Federal de Minas Gerais. Belo Horizonte, MG, Brasil}
\address[cefetmg]{Departamento de F\'\i sica e Matem\'atica, Centro Federal de Educa\c c\~ao Tecnol\'ogica de Minas Gerais, Belo Horizonte, MG, Brasil }

\begin{abstract}
According to the World Health Organization, dengue fever is the most important mosquito-borne disease of humans, and it is currently estimated that there may be 50 - 100 million yearly dengue infections worldwide. For the purpose to provide new techniques to public health  policies in course, we introduce a predictive non-linear population dynamics model to describe the population size of four stages of the development of \emph{Aedes aegypti}, having the coefficients set to be dependent on the rainfall index data. 
In spite of the population dynamics of the \emph{Ae. aegypti} be mainly ruled by the rainfall regime, most models are dedicated exclusively to effects of temperature and only few models are dedicated to influence of rainfall. Vector control actions are also implemented in many periods of the year in order to compare relative efficiency of public health policies.
The analysis of equilibrium and stability was performed. 
Field rainfall time series data from the City of Lavras (Minas Gerais, Brazil) was used for the model evaluation. 
The model was validated in a comparison with experimental mosquito abundance data acquired by field health agents. 
We evaluated and validated an entomological conjecture that claims that control actions should be performed during the dry season, instead of the common procedure adopted by vector control programs, in which those are mainly applied in the rainy season.
\end{abstract}


\end{frontmatter}

\section{Introduction}

\hspace{0.2in} According to the World Health Organization (WHO), dengue has become the most important human arbovirus disease. It is currently estimated that there may be 50 - 100 million yearly dengue infections worldwide, mainly in tropical and subtropical regions due to the disease environmental and climatic characteristics. Cases across the Americas, south-east Asia and western pacific, which exceeded 1.2 million cases in 2008, totalled 2.3 million in 2010 \citep{who2012denguefs117}. Several species of mosquitoes of the genus \emph{Aedes} are able to transmit the dengue virus, but \emph{Ae. aegypti} is the main dengue vector. 
In Brazil, the \emph{Ae. aegypti} mosquito is the only recognised dengue vector. Although the budget of government agencies for the prevention of the dengue has been growing every year worldwide, in most countries, those resources are often scarce. The strategic approach to prevent dengue infections recommended by WHO is the Integrated Vector Management (IVM) \citep{world2004global} and the ecologically well-established approach is to promote the environmental management of the vector population size \citep{ellis2009ecological}. However, most of those preventing actions remained almost unchanged for about one century  \citep{connor1923stegomyia, resende2013comparison}. Therefore, it is essential to increase the efficiency of prevention actions by restricting the \emph{Ae. aegypti} population size to acceptable levels in the environment.  

The \emph{environmental management} approach consists of \emph{vector surveillance} actions together with \emph{vector control actions}. \emph{Vector control actions} can be achieved via application of pesticides,  biological control and source removal. \emph{Vector surveillance} is based on sampling specific stages of mosquito development (eggs, larva/pupae or adult) to produce entomological indicators as a measure of the degree of infestation. The indicators for vector surveillance date from the first half of the 20th century and are based on collection of immature forms (larva/pupae) of the vectors \cite{breteau1954fievre}. 
The collection of adult forms, mainly gravid females, is currently conducted via the use of gravid adult traps. Several reports has demonstrated that gravid traps have the advantage of being strongly correlated with mosquito abundance and the number of dengue human cases \citep{kuno1995review, de2012dengue}, because females mosquitoes require human blood as a meal for oviposition and, consequently, are responsible for dengue transmission. Among adult traps, the sticky trap known as MosquiTRAP\textsuperscript{\textregistered} (Belo Horizonte, Brazil) is considered by the WHO as an effective method for surveillance and dengue prevention \cite{resende2012field}. The surveillance is performed by weekly inspect arrays of sticky traps distributed in dengue risk areas in municipalities. The collected data are electronically transmitted by mobile cellular systems to a centre for processing (MI Dengue - Intelligent Monitoring system of dengue vector, Ecovec\textsuperscript{\textregistered}, Belo Horizonte, Brazil) and is used to form the entomological indicator \emph{Mean Female Aedes Index} (MFAI), which corresponds to the number of females captured divided by the number of MosquiTRAP\textsuperscript{\textregistered} units installed \citep{eiras2009preliminary}. 

A long well-known connection between the vector development stages together with dengue transmission and meteorological factors \citep{focks1993dynamic,gubler1998resurgent} as well as climatological factors \citep{jetten1997potential, gubler2002global} causes dengue fever to be a seasonal disease. The increasing rainfall and high temperatures enhance the reproduction and survival of the dengue vector \citep{gubler2002global,chadee2005impact}. However, a set of urban and social conditions \citep{gubler1998resurgent,rigau1998dengue} enables the \emph{Ae. aegypti} to have sufficient vectorial density even in the seasonal dry period which favours the transmission of dengue  throughout the year \citep{kuno1995review, chadee2005impact}. Typically, the dengue vector control actions performed by vector control public programs are not permanent throughout the year and occur in periods in which the vector population increases under the influence of the rainfall regime \citep{LENZI2004}. Particularly in Brazil, as recommended by the Dengue Control National Program by the Ministry of Health \citep{pessanha2009avaliaccao}, the disease prevention efforts come into effect after the beginning of the rainfall season. A recurrent conjecture among entomologists claims \citep{chadee2005impact, pessanha2009avaliaccao} that the period of the actions of the dengue vector control public programs should be yearly set to the cold and dry seasons to reduce the vector population size and the number of infections, which will reduce the costs and social impact associated with the disease.

A number of entomological and epidemiological works have addressed the issue of the influence of meteorological variables upon the \emph{Ae. aegypti} population size and dengue transmission, mainly related to temperature effects \citep{Beserra2009, de2012dengue, Watts1987, lambrechts2011impact}. Some of those reports deal also with effect of rainfall \citep{de2012dengue}. There are mathematical and stochastic models developed to describe the influence of the temperature on dengue transmission \citep{pinho2010modelling, yang2008assessing} and on the dengue vector population dynamics \citep{ otero2006stochastic, lana2014seasonal}. However, in spite of the \emph{Ae. aegypti} population dynamics be mainly ruled by the rainfall regime, only few models are dedicated to its influence \citep{mondaini2012biomat}.

For the purpose to contribute to public health policies, a predictive entomological population dynamics model was designed to describe the population size of the stages of development of the \emph{Ae. aegypti} under the influence of rainfall time series data. The model is validated in comparison to experimental field data acquired by means of captures performed by an installed array of MosquiTRAP\textsuperscript{\textregistered} during the on course control action program. The control actions, also implemented in the model, were performed, and the effect of control in the dry and wet seasons was compared. In contrast to the usual approach of vector control programs, in which the control actions are performed during the rainy season, we analyse the efficiency advantage in performing the \emph{Ae. aegypti} mosquito control actions during the dry season to reduce the vector population size and the number of annual infections, and, in doing so, reducing the cost and social impact of dengue. In section \ref{model}, the model design is described, and the parameters are discussed. The analysis of equilibrium and stability is performed in section \ref{analysis}. The model is evaluated in section \ref{evaluation} using field rainfall data. In section \ref{validation}, a validation test is performed on the model by means of comparison with experimental capture field data. The above mentioned conjecture about efficiency of control performed in the dry season is verified in section \ref{verifying}. Conclusions are drawn in section \ref{conclusion}.

\section{The mathematical entomological model}
\label{model}
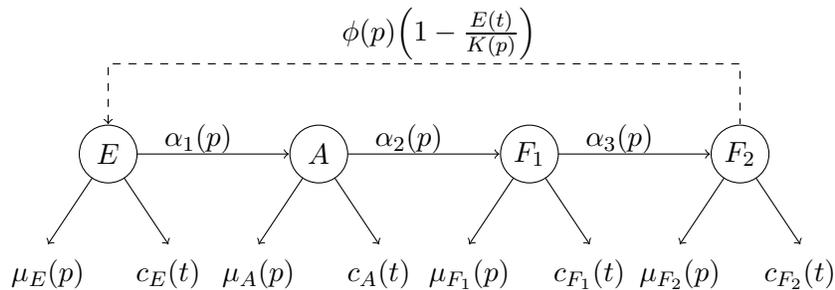
\begin{figure}[ht!]\label{esquemamodel}
     \centering
     \scaler{\input{modeldraw.tex}}{4.0cm} 
     \caption{\footnotesize{Diagram of the entomological model for the population dynamics of the stages of development of \emph{Aedes aegypti}.}}
     \label{fig1}
 \end{figure}

\hspace{0.2in} The present entomological model, describes the population dynamics of four development stages of \emph{Ae. aegypti} mosquito under the influence of a rainfall regime, having the coefficients set to be dependent on the weekly cumulative rainfall index in a entomologically plausible manner. The model, whose interactions are illustrated in figure \ref{esquemamodel}, is expressed by a system of non-linear differential equations (\ref{modelodengue3}). The dynamical variables, corresponding to populations are $\boldsymbol{X}(t)=(E(t), A(t), F_1(t), F_2(t))$. The population of eggs is represented by $E(t)$. Pupae and larvae have comparable short lifetimes. Therefore, they are considered together to constitute the aquatic population denoted by $A(t)$. The population $F_{1}(t)$ stands for the population of pre-bloodmeal females. The mated females, or post-bloodmeal females, which bite humans to feed their eggs is represented by $F_{2}(t)$.

  
The natural coefficients of the model are set to be parametrically dependent on the rainfall index $p$. The carrying capacity is denoted by $K$. The rate of oviposition performed by the post-bloodmeal females is symbolised by  $\phi$. The rate $\alpha_{1}$ corresponds to the development of eggs into an aquatic population. In turn, the rate in which individuals of the aquatic population develop into adult females in pre-bloodmeal stage is represented by $\alpha_{2}$.  The rate $\alpha_{3}$ stands for the development of pre-bloodmeal females into post-bloodmeal females. The rate of natural mortality of the populations are $\mu_{E}$,  $\mu_{A}$, $\mu_{F_{1}}$ and $\mu_{F_{2}}$, respectively, for eggs, aquatic stage, pre-bloodmeal and post-bloodmeal females. The non-natural rates $c_{E}$, $c_{A}$, $c_{F_{1}}$ and $c_{F_{2}}$, corresponding to control actions, are dependent on time t, as they are external parameters, and they can be turned on and off according to eventually applied public health policies. 
The only non-linear term $\phi \left (1 - \frac {E(t)} {K(p)} \right)F_2(t)$ \citep{yang2008assessing} encloses in the parentheses the term $\frac{E(t)}{K(p)}$ which mitigates the rate $\phi$ as the population of the stage $E(t)$ is sufficiently large if compared with the value of carrying capacity $K$, as the females avoid laying eggs in places containing a number previously laid eggs \citep{chadee1990egg}.


\begin{align}\label{modelodengue3}
&\frac{dE}{dt} = \phi(p) \left(1-\frac{E(t)}{K(p)}\right) F_{2}(t) - \alpha_{1}(p)E(t) - \mu_{E}(p)E(t) - c_{E}(t)E(t)\,,\nonumber\\ 
&\frac{dA}{dt} = \alpha_{1}(p)E(t) - \alpha_{2}(p) A(t) - \mu_{A}(p)A(t) - c_{A}(t)A(t)\,,\nonumber\\ 
&\frac{dF_{1}}{dt} = \alpha_{2}(p) A(t) - \alpha_{3}(p) F_{1}(t) - \mu_{F_{1}}(p)F_{1}(t) - c_{F_{1}}(t)F_{1}(t)\,,\\
&\frac{dF_{2}}{dt} = \alpha_{3}(p) F_{1}(t) - \mu_{F_{2}}(p)F_{2}(t) - c_{F_{2}}(t)F_{2}(t)\,,\nonumber\\[1ex]
&\phi,\,\, \alpha_{1},\,\,\alpha_{2},\,\,\alpha_{3},\,\,\mu_{E},\,\,\mu_{A},\,\,\mu_{F_{1}}, \mu_{F_{2}},\,\,K,\,\,c_{E},\,\,c_{A},\,\,c_{F_{1}},\,\,c_{F_{2}}\geq 0,\,\,\,\,\forall~p,t\in\mathbb{R}_+.\nonumber
\end{align}

We adopt a power-law expression for the parametric dependence of the model coefficients, generically represented by $\boldsymbol{\pi}=(\phi,\alpha_{1},\alpha_{2},\alpha_{3},\mu_{E},\mu_{A},\mu_{F1},\\ \mu_{F2},K)$, on rainfall index $p=p(t)$:

\begin{equation}
 \boldsymbol{\pi}(p) = \boldsymbol{\pi}_{0} + \frac{(\boldsymbol{\pi}_{1} - \boldsymbol{\pi}_{0})}{(p_{1} - p_{0})^r}(p - p_{0})^r \,.
 \end{equation}\label{parametrization}

The $p_{0}$ and $p_{1}$ values are consistent to average precipitation index values of tropical and sub\-tro\-pi\-cal zones, as well as $\boldsymbol{\pi}_{0}$ and $\boldsymbol{\pi}_{1}$, respectively, the associated reference values for the parameters in such climate or, incidentally, meteorological conditions. Given that the adaptability of \emph {Ae. aegypti} is formidable, the occurrence of small quantities of rainfall is considered sufficient to produce most effects of the population change. In contrast, when subjected to a large amount of precipitation, the response of the population to rainfall index variations is expected to be less sensitive. Thus, due to this entomologically plausible assumption, the power-law dependence was set as $0 < r < 1$ to represent this non-linear sensitivity on $p$. The cases $r=0$ and $r=1$ were used for constant and linear dependences, respectively.

\section{Analysis of equilibrium and stability}
\label{analysis}

\hspace{0.2in} As the model coefficients are set to be dependent on rainfall index $p$, which in turn varies with time, strictly speaking, the model is non-autonomous. However, the rainfall index is weekly accumulated so that the function $p=p(t)$ is constant by parts. Though at every epidemiological week, the coefficients are set to particular values, causing the phase portrait to change weekly, such a time scale is huge compared with the tick time evolution of the model step. Consequently, the model can be studied as a sequence of weekly updated autonomous models. Our analysis is carried out under such an assumption.

When imposing the equilibrium condition, ${\displaystyle\boldsymbol{X}^{\prime}(t)=0}$, the system (\ref{modelodengue3}) exhibits a trivial equilibrium point,
\begin{equation}\label{ptocritico3a}
P_a=(E(t)^*,A(t)^*,F_{1}(t)^*,F_{2}(t)^*)=(0,0,0,0)\,,
\end{equation}
and a non-trivial equilibrium point,
\begin{eqnarray}
\label{ptonaotrivial}
P_b&=&(E(t)^{**},A(t)^{**},F_{1}(t)^{**},F_{2}(t)^{**})~~= \nonumber
\\
&=&\left(K \left(1-\frac{1}{R_{0}}\right), \frac{\alpha_{1}}{(\alpha_{2} + \mu_{A} + c_{A})}E(t)^{**}, \right.  \\ 
& & \left. \qquad\qquad\qquad\frac{\alpha_{2}}{(\alpha_{3} + \mu_{F_{1}}+  + c_{F_{1}})}A(t)^{**}, \frac{\alpha_{3}(p)}{(\mu_{F_{2}} + c_{F_{2}})}F_{1}(t)^{**} \right)\,, \nonumber
\end{eqnarray}
where, $R_{0}$ is the \emph{basic reproduction number} and plays the role of a discriminant of the sign of the equilibrium point coordinates:

\begin{equation}\label{Rm}
R_{0}=\frac{\phi }{(\alpha_{1} + \mu_{E} + c_{E})}\frac{\alpha_{1}}{(\alpha_{2} + \mu_{A} + c_{A})}\frac{\alpha_{2}}{(\alpha_{3} + \mu_{F_{1}} + c_{F_{1}})}\frac{\alpha_{3}}{(\mu_{F_{2}} + c_{F_{2}})}\,.
\end{equation}

\begin{rem} \label{nonnegrates}
Given that all rates of the model (\ref{modelodengue3}) must be kept non-negative, it follows from (\ref{Rm}) that $R_{0}$ is defined so that $ R_{0}\geq 0$.
\end{rem}

In the model (\ref{modelodengue3}), both the equilibrium points and $R_{0}$ are parametrically dependent on the rainfall index $p$.  The case $R_{0}=1$ causes coinciding
$ P_b = P_a$ both as trivial equilibrium points.

\begin{rem} \label{nontrivcond}
For the model (\ref{modelodengue3}) to be ecologically plausible with equilibrium points associated to non-negative populations sizes of \emph{Ae. aegypti} development stages, from (\ref{ptonaotrivial}), the condition $R_{0}\geq 1$ must hold. Additionally, for the existence a non-trivial equilibrium point, $R_{0}>1$  is a necessary condition.
\end{rem}

To proceed with the stability study, local linear approximation in the neighbourhood of the equilibrium points (\ref{ptocritico3a}) and (\ref{ptonaotrivial}) was performed. The Jacobian matrix $B$ evaluated for the trivial equilibrium point (\ref{ptocritico3a}) reads as follows:

\begin{equation}
B_{P_a}=\left[
\begin{smallmatrix}
   - (\alpha_{1} + \mu_{E} + c_{E}) & 0 & 0  & \phi \\
    \alpha_{1} & - (\alpha_{2} + \mu_{A} + c_{A}) & 0 & 0\\
   0 & \alpha_{2} & - (\alpha_{3} + \mu_{F_{1}} + c_{F_{1}}) & 0\\
   0 & 0 & \alpha_{3} & - (\mu_{F_{2}} + c_{F_{2}})
\end{smallmatrix}
\right] \label{jacobianP0}.
\end{equation}

The characteristic polynomial of order 4 in $\lambda$ have the form
\begin{equation}
\label{polyn}
a_{0}\lambda^4 + a_{1}\lambda^3 + a_{2}\lambda^2 + a_{3}\lambda + a_{4} = 0\,,
\end{equation}
where the coefficients are as follows:

\begin{align}
\label{coeffp0}
a_{0}^* &=  1 \nonumber\\\nonumber\\
a_{1}^* &=  (\mu_{F_{2}} + c_{F_{2}})+ (\alpha_{1} + \mu_{E} + c_{E}) + (\alpha_{2} + \mu_{A} + c_{A}) +  (\alpha_{3} + \mu_{F_{1}} + c_{F_{1}})>0, \nonumber\\\nonumber\\
a_{2}^* &= (\mu_{F_{2}} + c_{F_{2}})[(\alpha_{1} + \mu_{E} + c_{E}) + (\alpha_{2} + \mu_{A} + c_{A}) + (\alpha_{3} + \mu_{F_{1}} + c_{F_{1}})] +\nonumber\\
&+ (\alpha_{2} + \mu_{A} + c_{A})[(\alpha_{1} + \mu_{E} + c_{E}) +  (\alpha_{3} + \mu_{F_{1}} + c_{F_{1}})] \nonumber\\
&+ (\alpha_{3} + \mu_{F_{1}} + c_{F_{1}})(\alpha_{1} + \mu_{E} + c_{E})>0,  \\ \nonumber\\
a_{3}^* &= (\alpha_{2} + \mu_{A} + c_{A})(\alpha_{1} + \mu_{E} + c_{E})[(\mu_{F_{2}} + c_{F_{2}}) + (\alpha_{3} + \mu_{F_{1}} + c_{F_{1}})] \nonumber\\
&+ (\alpha_{3} + \mu_{F_{1}} + c_{F_{1}})(\mu_{F_{2}} + c_{F_{2}})[(\alpha_{2} + \mu_{A} + c_{A}) + (\alpha_{1} + \mu_{E} + c_{E})] > 0,\nonumber\\\nonumber\\
a_{4}^* &= (\alpha_{3} + \mu_{F_{1}} + c_{F_{1}})(\mu_{F_{2}} + c_{F_{2}})(\alpha_{2} + \mu_{A} + c_{A})(\alpha_{1} + \mu_{E} + c_{E})(1 - R_{0}). \nonumber
\end{align}

\begin{thm} \label{thm1}
If the conditions of non-negativity ($R_{0}\geq 1$) as well as of non-triviality ($R_{0} \neq 1$) of the equilibrium points are applied so that $R_{0}>1$, then the trivial equilibrium point $P_a$ of the model (\ref{modelodengue3}) is unstable. 
\end{thm}

\begin{prf}
From the Descartes's rule of signs \citep{wiggins2003introduction}, if the total number of sign changes in the sequence of coefficients $(a_{4}^*,a_{3}^*, a_{2}^*, a_{1}^*, a_{0}^* )$ of the polynomial (\ref{polyn}) coefficients is $k$, then the number of positive real roots is given by one of the following cases: $k-n\geq 0$, $n=0,2,4,\cdots,N$. Where $N=k$, for even $k$ or $N=k-1$ for odd $k$. Since, only $a_{4}^*<0$ is negative as $R_0>1$ in (\ref{coeffp0}), there is only a change of sign in the sequence of polynomial coefficients, so the only possible case is $k=1$ and $n=0$. Thus, there exists one positive real root for the characteristic polynomial. This is sufficient to state that $P_a$ is an unstable equilibrium point.
\hfill$\blacksquare$
\end{prf}


The jacobian matrix $B$, evaluated in equilibrium non-trivial point (\ref{ptonaotrivial}), is given by
\begin{equation}
B_{P_b}=\left[
\begin{smallmatrix}
-\phi(\alpha_{1}+\mu_{E}+c_{E})&0&0 &\frac{\phi}{R_{0}} \\
    \alpha_{1}&-(\alpha_{2}+\mu_{A} + c_{A})&0&0\\
   0&\alpha_{2}&-(\alpha_{3} + \mu_{F_{1}}+c_{F_{1}})&0\\
   0&0&\alpha_{3}&-(\mu_{F_{2}}+c_{F_{2}})
\end{smallmatrix} 
\right] \label{jacobianP1}
\end{equation}

For the non-trivial equilibrium point (\ref{ptonaotrivial}), the coefficients of the characteristic polynomial (\ref{polyn}) are:

\begin{align}
\label{coeffp1}
a_{0}^{**} &=  1 \nonumber\\ \nonumber\\
a_{1}^{**} &= (\mu_{F_{2}} + c_{F_{2}})+ (\alpha_{2} + \mu_{A} + c_{A}) +  (\alpha_{3} + \mu_{F_{1}} + c_{F_{1}}) +\nonumber\\ 
&+ \frac{\phi \alpha_{1} \alpha_{2} \alpha_{3}}{(\alpha_{3}+\mu_{F_{1}}+c_{F_{1}})(\mu_{F_{2}}+c_{F_{2}})(\alpha_{2} + \mu_{A} + c_{A})}>0, \nonumber\\\nonumber\\
a_{2}^{**} &= (\mu_{F_{2}} + c_{F_{2}})[(\alpha_{2} + \mu_{A} + c_{A}) + (\alpha_{3} + \mu_{F_{1}} + c_{F_{1}})] +\nonumber\\ 
&+\left[(\alpha_{2} + \mu_{A} + c_{A})(\alpha_{3} + \mu_{F_{1}} + c_{F_{1}})+\frac{\phi \alpha_{1} \alpha_{2} \alpha_{3}}{(\alpha_{3} +\mu_{F_{1}}+c_{F_{1}})(\mu_{F_{2}}+c_{F_{2}})}\right] + \nonumber\\ 
&+\left[\frac{\phi \alpha_{1} \alpha_{2} \alpha_{3}}{(\mu_{F_{2}}+c_{F_{2}})(\alpha_{2} + \mu_{A} + c_{A})} + \frac{\phi \alpha_{1} \alpha_{2} \alpha_{3}}{(\alpha_{3}+\mu_{F_{1}}+c_{F_{1}})(\alpha_{2} + \mu_{A} + c_{A})}\right]> 0, \nonumber\\ \nonumber\\
a_{3}^{**} &=  \left[\frac{\phi \alpha_{1} \alpha_{2} \alpha_{3}}{(\mu_{F_{2}} + c_{F_{2}})} + \frac{\phi \alpha_{1} \alpha_{2} \alpha_{3}}{(\alpha_{3} + \mu_{F_{1}} + c_{F_{1}})} + \frac{\phi \alpha_{1} \alpha_{2} \alpha_{3}}{(\alpha_{2} + \mu_{A} + c_{A})} \right] +  \\
&+ (\alpha_{3} + \mu_{F_{1}} + c_{F_{1}})(\mu_{F_{2}} + c_{F_{2}})(\alpha_{2} + \mu_{A} + c_{A})> 0, \nonumber\\ \nonumber\\
a_{4}^{**} &= \phi \alpha_{1} \alpha_{2} \alpha_{3}   \left(1 - \frac{1}{R_{0}}\right). \nonumber
\end{align}

\begin{thm} \label{thm2}
If the conditions of non-negativity ($R_{0}\geq 1$) as well as of non-triviality ($R_{0} \neq 1$) of the equilibrium points are applied so that $R_{0}>1$, then the equilibrium point $P_b$ of the model (\ref{modelodengue3}) is asymptotically stable.
\end{thm}

\begin{prf}
If the condition of non-negativity of the equilibrium population together with the condition of non-triviality are imposed, $R_{0}>1$, the polynomial coefficients (\ref{coeffp0}) are so that $a_1^*>0$, $a_2^*>0$, $a_3^*>0$ and $a_4^*>0$. In accord to the Routh-Hurwitz test \citep{wiggins2003introduction}, all the roots of the polynomial (of degree four, in this case) have strictly negative real parts if and only if the following conditions are satisfied: $ a_{1}> 0 $, $ a_{4}> 0 $, $  a_{1} a_{2}- a_{0}a_{3}> 0 $ and $ a_{1} a_{2} a_{3}> a_{0}a_{3} ^{2} + a_{1}^{2}a_{4} $. Since the whole conditions are satisfied, $P_b$ is an asymptotically stable equilibrium point of the model (\ref{modelodengue3}).
\hfill$\blacksquare$
\end{prf}

\begin{figure}[h!]
\centering
  \includegraphics[width=0.7\textwidth]{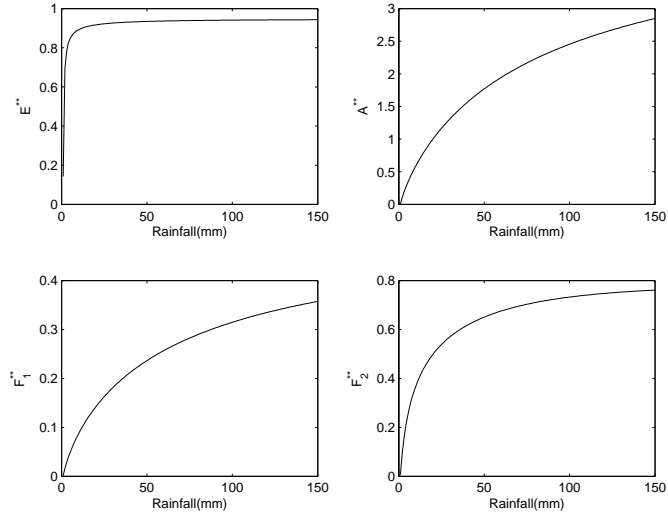}
\caption{\footnotesize{Curve described in the state space by the non-trivial equilibrium point~(\ref{ptonaotrivial}) as the rainfall index $p$ varies weekly in time in the range of $[0,150]$ mm/week.}}
\label{parametz}
\end{figure}

As $p$ varies, the coordinates of $P_b$ describe a curve in the state space. The parametric functions of the components of such curve is depicted in figure (\ref{parametz}).

\section{Evaluating the model}
\label{evaluation}

\hspace{0.2in} Model (\ref{modelodengue3}) was numerically implemented with the Runge-Kutta fourth order method in MATLAB R2009b (MathWorks Inc., mathwork.com, Natick, MA, USA). The reference values of the rainfall parameters were fixed so that $p_{0}=0$, corresponding to dry weather, and $p_{1}$, consistent with the lower limit of the mean weekly rainfall of tropical rainforests (~$1800$~mm/year, ~$34.62$~mm/week) which is supposed to be close to ideal conditions of rainfall for the development of \emph{Ae. aegypti}. Also, in so doing, the amplitude of the rainfall index variation of various ecosystem climates are well represented in the range of reference values. The maximum and minimum values of the entomological coefficients adopted are shown in table (\ref{tab:0}). Some of those values were obtained from \citet{Ferreira2003a} and \citet{yang2008assessing}, and others were estimated by entomologist authors. The used values of $r$ were $r=0.85$ for $\phi,\alpha_{1},\alpha_{2},\alpha_{3})$, $r=1$ for $K,\mu_{F1},\mu_{F2}$, and $r=0$ for $\mu_{A},\mu_{E}$. In the model implementation, the condition $R_{0}>1$ was verified for each one of the values of the rainfall index $p$. 


\begin{table}[h!]
\caption{\footnotesize{Range of parameters of the model, in which the  minimum and maximum values correspond to week rainfall index $p_{min}=0\,\mbox{mm/weak}$ and $p_{max}=34.62\,\mbox{mm/weak}$, respectively.}}
\centering
\label{tab:0}
\begin{tabular}{cccc}
  $Rate$ & $Range$ & $Rate$ & $Range$ \\
   \hline 
  $\phi$ & $0.560~-~11.2$ & $K$  & $1.00~-~1.00$ \\
  \hline
  $\alpha_{1}$ & $0.0100~-~0.500$ & $\alpha_{2}$ & $0.0600~-~0.160$\\
  \hline
  $\mu_{E}$ & $0.0100~-~0.0100$ & $c_{E}$ & $0.300~-~0.300$ \\
  \hline
  $\mu_{A}$ & $0.164~-~0.164$ & $c_{A}$ & $0.300~-~0.300$ \\
  \hline
  $\mu_{F_{1}}$  & $0.0430~-~0.170$ & $\mu_{F_{2}}$  & $0.0570~-~0.333$\\
  \hline
  $c_{F_{1}} = c_{F_{2}}$  & $0.000~-~0.000$ & $\alpha_{3}$ & $0.333~-~1.00$  \\
  \hline
\end{tabular}
\end{table}

The rainfall index data used as input for the set of linear dependent parameters was the weekly accumulated rainfall index of the City of Lavras (Minas Gerais, Brazil), which were obtained from National Institute for Space Research (INPE, Brazil). The data refers to the epidemiological weeks 9 to 52 of the year 2009 and to the epidemiological weeks 1 to 46 of the year 2010. The carrying capacity was set to one, $K=1$, causing the populations sizes to vary as fractions of the unit.

\begin{figure*}[h!]
\centering
  \includegraphics[width=0.7\textwidth]{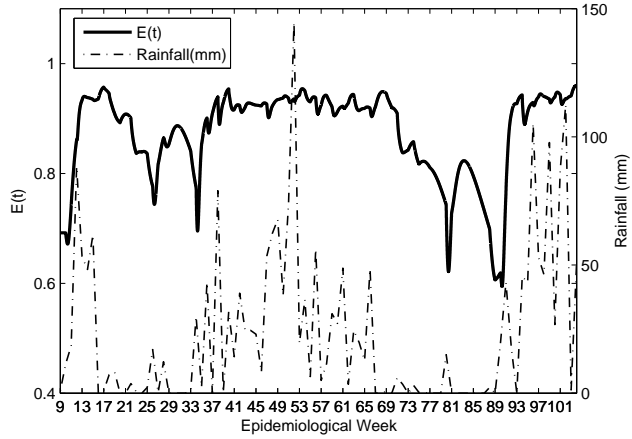}
\caption{\footnotesize{Modelled time dynamics of the population size of \emph{Aedes aegypti} eggs $E(t)$ in comparison with the rainfall data.}}
\label{popE}
\end{figure*}

\begin{figure*}[h!]
\centering
  \includegraphics[width=0.7\textwidth]{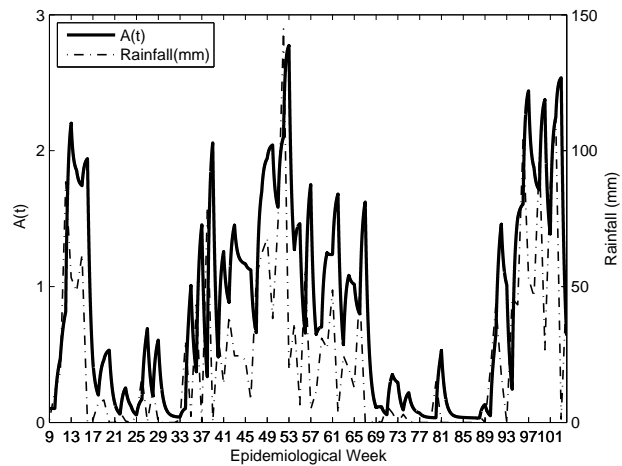}
\caption{\footnotesize{Modelled time dynamics of the population size of \emph{Aedes aegypti} aquatic forms $A(t)$ in comparison with the rainfall data.}}
\label{popA}
\end{figure*}

\begin{figure*}[h!]
\centering
  \includegraphics[width=0.7\textwidth]{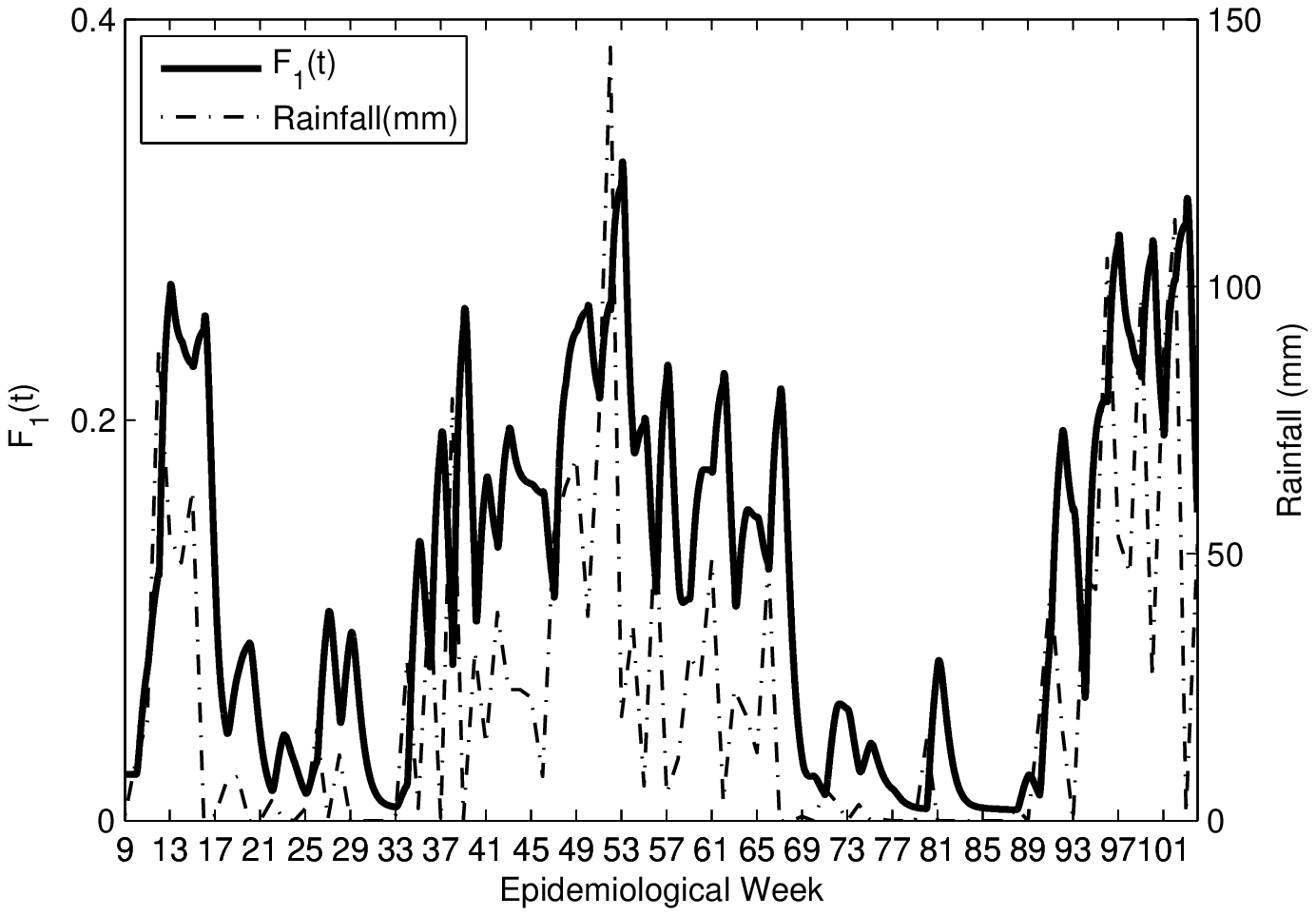}
\caption{{Modelled time dynamics of the population size of \emph{Aedes aegypti} pre-bloodmeal females $F_1(t)$ in comparison with the rainfall data.}}
\label{popF1}
\end{figure*}

\begin{figure*}[h!]
\centering
  \includegraphics[width=0.7\textwidth]{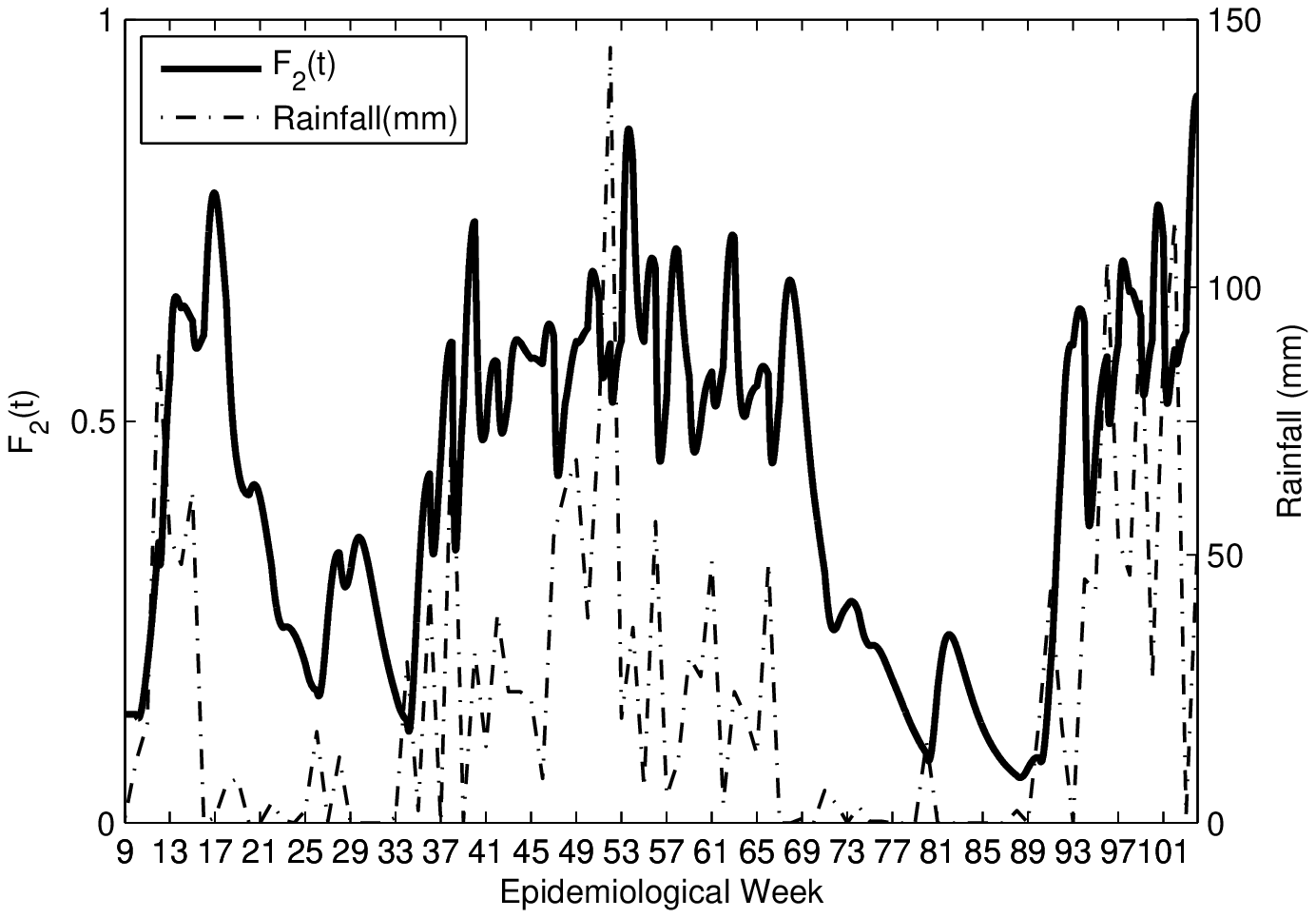}
\caption{\footnotesize{Modelled time dynamics of the population size of \emph{Aedes aegypti} post-bloodmeal females $F_2(t)$ in comparison with the rainfall data.}}
\label{popF2}
\end{figure*}

Figures \ref{popE} to \ref{popF2} illustrate the evolution of the population sizes of the stages of development of \emph{Ae. aegypti} without controls compared with the rainfall index for the City of Lavras (Minas Gerais, Brazil) over the study period.

It can be noticed that the time evolution of the population sizes of the development stages, $E(t)$, $A(t)$ and $F_1(t)$ have peaks reacting to the peaks of the rainfall index with a certain delay of approximately one week, and population $F_2(t)$ exhibits a delay of approximately two weeks. The patterns of the results are very similar to the analogous experimental data reported by \citet{de2012dengue}.


\section{Model validation}
\label{validation}

\hspace{0.2in} A validation test was performed comparing the modelled population of post-bloodmeal females $F_2(t)$ with the experimental field data of sampled mosquito abundance obtained by trapping \emph{Ae. aegypti} adult females. The data series for the City of Lavras (Minas Gerais, Brazil) was provided by Ecovec\textsuperscript{\textregistered} and was acquired on the course of the control action program by public health agents by means of captures performed by an installed array of MosquiTRAP\textsuperscript{\textregistered}. The capture data was extracted to form the MFAI indicator time series over the same period of 2009-2010 of the rainfall data used in the model execution.

\begin{figure*}[h!]
\centering
  \includegraphics[width=0.7\textwidth]{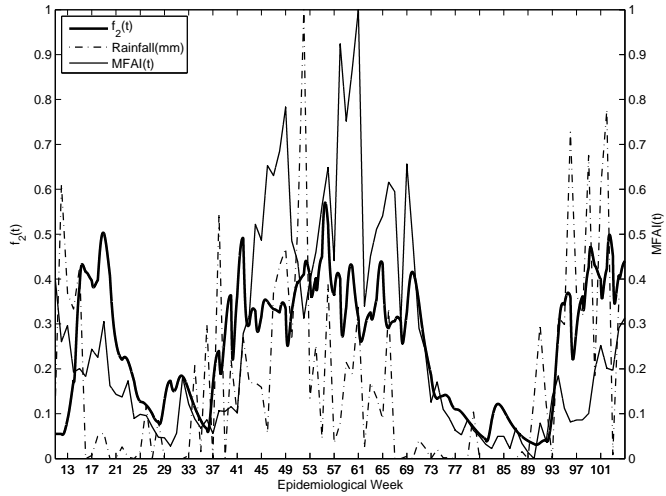}
\caption{\footnotesize{Comparison between the modelled $F_2(t)$, translated and normalised population to form $f_2(t)$ with the experimental field data of captures provided by the MFAI entomological indicator. The rainfall data is also included normalised to provide a metric for time comparison.}}
\label{F2IMFA}
\end{figure*}

To compare the modelled post-bloodmeal females population $F_2(t)$ with the MFAI data series, the $F_2(t)$ population was translated in time to best coincide with the peaks by means of a cross-correlation function. After the translation, $F_2(t)$ was normalized via optimisation with summed least-square cost function to form $f_2(t)$. This is sensible and not arbitrary procedure because the comparison is performed between distinct quantities: a (theoretical) population $F_2(t)$ and a sample of the natural population by means of the MFAI indicator and, because there is an expected time lag between the peaks of the natural field population size and the peaks of the experimental sample capture data of the MFAI. This is due to the delay caused by the accumulated weekly based survey. The  figure \ref{F2IMFA}  depicts the comparison between the experimental indicator of captures MFAI acquaired by adult traps against the translated and normalized theoretical female population size, $f_2(t)$.

Taking into account that the MFAI indicator is essentially a sample of the female population size that may be or not proportional to the overall natural population size, it can be notice in figure \ref{F2IMFA} that there are reasonable coincidences in the position of peaks of $f_2(t)$ and the MFAI. Furthermore, in the graph, there are three periods in which a large amount of captures occur, indicating high MFAI values. These accumulations correspond to the rainy seasons in which experimental data was consistently followed by the modelled $f_2(t)$ data, though with some discrepancies. In the central region with accumulated captures, the model does not reach the experimental value of IMFA indicator; while in the accumulation regions of both sides, the model exceeds the experimental data. In both cases the model exhibits a certain degree of saturation as the peaks have similar height.

\begin{figure*}[h!]
\centering
  \includegraphics[width=0.7\textwidth]{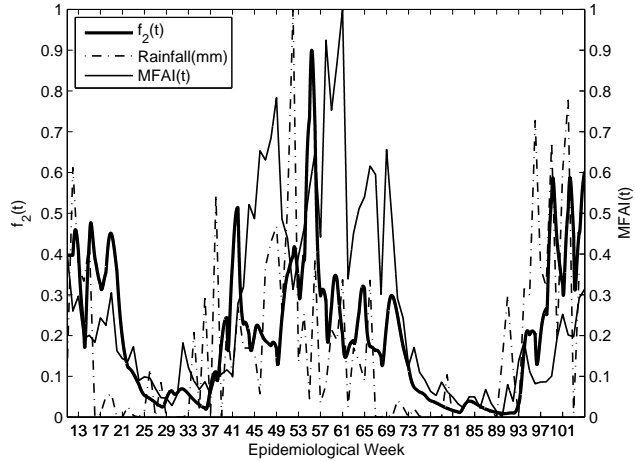}
\caption{\footnotesize{Comparison between the modelled $F_2(t)$, translated and normalised population to form $f_2(t)$ with the experimental field data of captures provided by the MFAI entomological indicator with varying carrying capacity $K(p)$. The rainfall data is also included normalised to provide a metric for time comparison.}}
\label{F2IMFAK}
\end{figure*}

We further investigate the dependence of the carrying capacity $K$ on the rainfall index $p$ to simulate the impact on post-bloodmeal females $F_2(t)$ population of short-time temporary breeding sites which takes place in weeks of high rainfall index. The dependence of the function $K(p)$ was set as the equation (\ref{parametrization}) with $r=0.85$ and the result is shown in the figure \ref{F2IMFAK}. The result resembles that of the case $K=1$ with less saturation in rainy seasons.

\section{Verifying the conjecture}
\label{verifying}

\hspace{0.2in} According to the afore mentioned conjecture, the period of the actions of the dengue vector control public programs should be advanced to the cold and dry seasons to reduce the vector population size and reduce the number of annual dengue infections \citep{chadee2005impact, pessanha2009avaliaccao}. As the model was qualitatively validated in the last section, verification of the conjecture is now possible. For that purpose, the control actions were modelled over the populations $E(t)$ and $A(t)$ to simulate the approach adopted by the dengue vector control public programs, which consists in the removal of breeding sites \citep{rigau1998dengue}. The effectiveness of this type of control was estimated and set to be constant over a specified week to remove approximately $30 \%$ of the average week population. The carying capacity was set to $K=1$. We focused our analysis on the population $F_2(t)$, as the gravid females are the form responsible for the infection of dengue fever and also because this population is captured by adult traps designed to resemble a breeding site.

The model was evaluated in three different conditions, conserving the same rainfall data series. Firstly, a set of model evaluations was performed with vector control theoretically applied over individual low rainfall index weeks (LRW). Analogously, it was evaluated with a vector control theoretically performed over a set of individual high rainfall index weeks (HRW). Finally, the model evaluation without any control weeks (WCW). The results of the above simulations of strategies of control actions were organised in pairs: (1) control over a week in dry season with the case without any control weeks (LRW, WCW); (2) control in a wet season with the case without any control weeks (HRW, WCW).

\begin{defn} \label{defreldif}
In order to compare the relative effectiveness of the control strategies, we define the relative differences indices $M(t)$ and $N(t)$ between the case without any control (WCW) and, respectively, the cases of control performed in a week of the dry season (LRW) and a week of the rainy season (HRW), such that:
\begin{equation}\label{N}
\!\!M(t) = \frac{F_{2}^{LRW}(t) - F_{2}^{WCW}(t)}{F_{2}^{WCW}(t)}~~~~\text{and}~~~~N(t) = \frac{F_{2}^{HRW}(t) - F_{2}^{WCW}(t)}{F_{2}^{WCW}(t)}.
\end{equation}
\end{defn}

\begin{figure*}[h!]
\centering
  \includegraphics[width=0.6\textwidth]{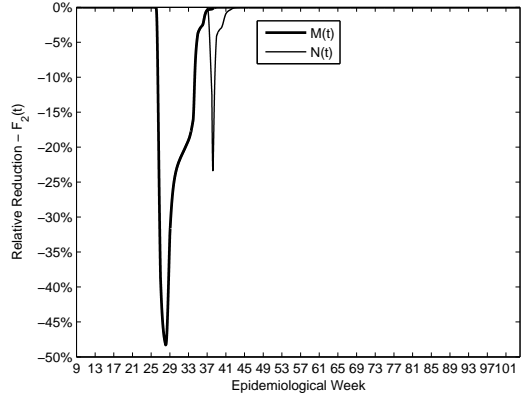}
\caption{\footnotesize{ Comparison of the relative differences between vector control performed in a low rainfall week and the control performed in a high rainfall week theoretically evaluated for the City of Lavras (Minas Gerais, Brazil).}}
\label{reducao2738}
\end{figure*}

The figure \ref{reducao2738} illustrates the comparison between the relative efficiency of the control performed with LRW number 27 (LRW-27) and HRW number 38 (HRW-38). The case of the control performed in the dry season (week 27 - LRW) resulted in the a depletion of the theoretical female population $F_2(t)$ at approximately two times the levels of depletion achieved by the control carried out in the wet season (week 38 - HRW).

\begin{defn} \label{defints}
The depth of the peaks of population depletions (figure \ref{reducao2738}) is as important as the time latency of the depletion. Thus, the best metric to account for the effect of population reduction is the area comprised by the peaks and the horizontal axis. Thus, the following indeces based on integrals was defined:
\begin{equation}\label{I}
I_{LRW} = \int_{I} M(t) dt~~~~~\text{and}~~~~I_{HRW} =\int_{I} N(t) dt.
\end{equation}\end{defn}

\begin{defn} \label{defrpercdif}
We define the relative percentage difference between the population depletion areas $I_{LRW}$ and $I_{HRW}$ indices to provide a comparison between the effectiveness between the control actions performed in the wet season against that performed in the dry season.
\begin{equation}\label{integral}
\Delta = \frac{I_{LRW}-I_{HRW}} {I_{HRW}} \times 100\%.
\end{equation}
\end{defn}

The table (\ref{tab:1}) shows the relative percentage differences  $\Delta$ of the areas of the peak depletions associated with the case of control implemented in pairs of weeks (LRW, HRW).

\begin{center}
\begin{table}[h]
\caption{\footnotesize{Relative percentage difference between the area of the peaks of population depletion for the pair of cases of control theoretically performed in a dry season week and in a wet season week for the City of Lavras (Minas Gerais, Brazil).}}
\centering

\begin{tabular}{cccc}
\hline
 \backslashbox{LRW}{HRW} & 38 & 47 & 51\\
\hline
20 &  643.0\%&  452.5\% &  290.9\%\\
 \hline
27 &  809.1\% &  576.1\% &  378.3\%\\
 \hline
 29 &  792.3\% &  563.6\% &  369.5\%\\
\hline
\end{tabular}\label{tab:1}
\end{table}
\end{center}

Analyzing table (\ref{tab:1}), we found that the control when performed in epidemiological weeks of low rainfall index appears to be more effective, causing relative depletion to be deeper and longer such that the areas of depletion range from 290.9$\%$ to 809.1$\%$. The control in LRW was found to be significantly advantageous compared with the control in HRW for the pairs of weeks (week 20 - LRW, week 38 - HRW), (week 27 - LRW, week 38 - HRW), (week 27 - LRW, week 47 - HRW), (week 29 - LRW, week 38 - HRW), and (week 29 - LRW, week 47 - HRW) ranging from 563.6$\%$ to 809.1$\%$.

\section{Conclusion}
\label{conclusion}
\hspace{0.2in} The validation of the model via comparison of the field data of captures with the modelled po\-pu\-la\-tion of females indicated reasonable similarities in position of the peaks and in the position of high infestation areas, but there were also differences. In some rainy seasons, the model population becomes greater than the capture indicator MFAI and in others, the opposite occurs. Such discrepancies could have contributions of some independent factors. The first is the temperature dependence of the populations. The present model is not designed to consider the effects of the temperature, though it is well established that temperature affects the rates of development stages of \emph{Ae. aegypti}. Secondly, when comparing those curves, the underlying assumption is that the array of adult traps provides a fair sampling of the field population. This assumption may not be completely true, as the efficiency of the adult trap could be impacted by the change of behaviour of the mosquitoes as the rainfall index varies, resulting in intrinsic dependence of the sampling with the precipitation. In other words, the sampling method of capturing using adult traps, though positively correlated to the natural population size dynamics, does not correspond to the same phenomenon as the number of captures by the adult traps could itself vary with rainfall. From those considerations, it can be said that the model has dynamic population behaviour similar to that expected for the natural \emph{Ae. aegypti} population. Another question that may have consequence in the results concerns the purpose of the field inspections that, though they are weekly conducted, they are designed to obtain accurately average of captures over three consecutive weeks and, when the rainfall index is high, the efficiency on the monitoring is low, since, due the difficulties posed by the rainfall on field agents labour, some traps are left to be inspected in the following week.  

Although the power law dependence of the coefficients with rainfall is entomologically plausible to be concave down ($r<1$) in the sense that, the production of mosquitoes is more sensitive to a small amount of precipitation and less sensitive to a large amount, it is not expected that all the coefficients of the model exhibit the same power law dependence. For instance, when the same power law is used in the whole set of coefficients, the value $r=0.85$ provided a best fit between MFAI and gravid female theoretical population $f_2(t)$, but the general case of different $r$-values for distinct coefficients is likely to produce better fit.

The modelled control actions produced up to 809.06$\%$ more effective population reduction and, on average, 542.4$\%$ more effective population reduction, if performed over an epidemiological week with low rainfall index as compared with control performed over an epidemiological week with a high rainfall index. This is important since the dry eggs are kept viable for months and a number of eggs hatching in the beginning of the rainy season were stored during the dry season. Thus, under the assumption that the model is sufficiently similar to the behaviour of \emph{Ae. aegypti} population, the conjecture about the advantage on effectiveness of the control performed in the dry season was favourably verified.

Future works should include along with the rainfall, the effect of temperature and humidity in the entomological parameters of the model (\ref{modelodengue3}). Optimization methods can be used to provide refinement of the dependence of the entomological parameters of the model with meteorological parameters. Similarly, the choice of power law dependence of the parameters of the model (\ref{modelodengue3}) with rainfall index can be improved by optimization processes to suitable non-linear dependencies, with different laws for the various climate parameters. Afterwards, the time intervals at which control can be performed could also be an object of study with the goal of obtain the best form and period to perform control by public health policies.

\section{Acknowledgments}
J. L. Acebal would like to express his gratitude to Atac\'ilio C. Alves who first brought him the problem concerning dengue vector. The research for this paper was financially supported by the Coordination for Enhancement of Higher Education Personnel, CAPES, Brazil the Federal Centre for Technological Education of Minas Gerais State, CEFETMG, Brazil and the National Council for Research and Development, CNPq, Brazil.

\bibliography{AedesDengue201306}

\end{document}

%% file: modeldraw.tex
\begin{tikzpicture}[scale=0.8]
    \clip (-1.5,-2) rectangle (13,3);
    
    \draw[->,draw=black,]  (0.5,0.5)--++(3,0);
    \node at (2.0,0.75) {$\alpha_{1}(p)$};

    \draw[->,draw=black]  (4.0,0.5)--++(3,0);
    \node at (5.5,0.75) {$\alpha_{2}(p)$};

    \draw[->,draw=black]  (7.5,0.5)--++(3,0);
    \node at (9.0,0.75) {$\alpha_{3}(p)$};

    \draw[<-,draw=black,dashed] (0.5,1)--++(0,1)--++(10.5,0)--++(0,-1);
    \node at (6,2.5) {$\phi(p) \Big(1 - \frac{E(t)}{K(p)} \Big)$};

    \draw[->] (0.5,0.5)--(1.5,-1);
    \draw[->] (0.5,0.5)--(-0.5,-1);
    \node at (-0.5,-1.5) {$\mu_E(p)$};
    \node at (1.5,-1.5) {$c_E(t)$};
    \draw[fill=white] (0.5,0.5) circle (0.48cm);
    \node at (0.5,0.5) {\bf$E$};

    \draw[->] (4.0,0.5)--(5.0,-1);
    \draw[->] (4.0,0.5)--(3.0,-1);
    \node at (3.0,-1.5) {$\mu_A(p)$};
    \node at (5.0,-1.5) {$c_A(t)$};
    \draw[fill=white] (4.0,0.5) circle (0.48cm);
    \node at (4.0,0.5) {\bf$A$};

    \draw[->] (7.5,0.5)--(8.5,-1);
    \draw[->] (7.5,0.5)--(6.5,-1);
    \node at (6.5,-1.5) {$\mu_{F_1}(p)$};
    \node at (8.5,-1.5) {$c_{F_1}(t)$};
    \draw[fill=white] (7.5,0.5) circle (0.48cm);
    \node at (7.5,0.5) {\bf$F_1$};

    \draw[->] (11,0.5)--(12.0,-1);
    \draw[->] (11,0.5)--(10.0,-1);
    \node at (10.0,-1.5) {$\mu_{{F_2}}(p)$};
    \node at (12.0,-1.5) {$c_{F_2}(t)$};
    \draw[fill=white] (11,0.5) circle (0.48cm);
    \node at (11,0.5) {\bf$F_2$};
  \end{tikzpicture}